\newif\iffiginclude
\figincludefalse        
\iffiginclude
  \documentstyle[12pt,psfig]{article}
\else
  \documentstyle[12pt]{article}
\fi
\def\lsim{\lower3pt\hbox{$\buildrel<\over\sim$}}
\date{December 1993}

\def\to{\rightarrow}
\begin{document}
\title{Comparison of Diagnostic $Z'$ Physics  at  Future  $pp$ and $e^+e^-$
Colliders\footnote{Contribution to
the Proceedings of the International Europhysics
Conference on High Energy
Physics, Marseille, July  21-28, 1993.}
}
\author{M. Cveti\v c\\
Department of Physics,\\
University of Pennsylvania,\\
Philadelphia, PA 19104--6396, USA\\
\ \ \\
UPR-597-T\\
 }

\maketitle
\vskip5mm
\begin{abstract}
{\rightskip=1.5pc\leftskip=1.5pc
We present recent developments in the diagnostic study of
heavy gauge bosons at future $pp$ (CERN LHC) and $e^+e^-$  (NLC)
colliders  with the  emphasis  on the model independent
determination of gauge couplings of $Z'$ to quarks and leptons.
The  analysis reflects a complementary diagnostic power of
the LHC  and the NLC (c.m. energy 500  GeV, integrated luminosity 20 fb$^{-1}
$).
At the NLC   for $M_{Z'} \sim 1 $  TeV  {\it all } the  quark and  lepton
charges
can be determined to around $10-20 \%$, provided heavy flavor tagging
and   longitudinal polarization of the  electron beam is available. At the
LHC primarily the magnitude of three (out of four) normalized
couplings can be determined, however  typical uncertainties are   by
a factor $\sim  2$
smaller than at the NLC. }
\end{abstract}

\twocolumn
{\bf\noindent Introduction}
\vskip2.5mm
 If heavy gauge bosons $Z'$'s
turn out to have a mass up to around 5  TeV,
future hadron colliders, {\it e.g.}, the large hadron  collider
 (LHC) at CERN, would be an ideal place to discover them.
An immediate need after $Z'$  discovery  would be to
learn more about its  properties.\footnote{For a recent review of
$Z'$ physics at future hadron colliders see  Ref. ~\cite{ACLII} .}
 In particular,  a determination of  $Z'$ couplings to quarks and
leptons is useful.  By now a series of such probes was proposed, allowing
for a model independent  determination of quark and lepton charges provided
$M_{Z'}\lsim 2$ TeV.

On the other hand, future $e^+e^-$ colliders
 with large enough  c.m. energy, {\it e.g.}, $\sqrt{s} =2$
TeV, could provide a clean way to discover and study the properties of $Z'$'s.
A more likely possibility, however, is the next linear collider (NLC)
 with $\sqrt{s}=500$ GeV.
There, due to the interference effects of the $Z'$ propagator with the photon
and
$Z$  propagator, the   probes with the  two--fermion  final states
yield a complementary   information on
the existence of a  $Z'$.  An extensive study \cite{DLRSV,HRV}\ showed that
effects of $Z'$'s would be observable at the NLC for  a large class of models
with $M_{Z'}$ up to  $1-3$ TeV.  In particular,
sensitivity of the NLC to discriminate \cite{DLRSV}\   between specific
classes of
 extended electroweak models, {\it e.g.},
different $E_6$ motivated models described by a parameter $\cos \beta$ (the
mixing between the $Z_\chi$ and the $Z_\psi$ defined below) or
left-right symmetric models parameterized by the ratio $\kappa=g_R/g_L$ for
the $SU(2)_{L,R}$ gauge coupling constants $g_{L,R}$,
was explored. Most recently, a model independent determination of $Z'$
couplings to quarks and leptons and a  comparison  with the LHC colliders was
performed in Ref. \cite{ACLIII} .\footnote{ For a
related work on the bounds for leptonic gauge couplings see  Ref. \cite{LEIKE}\
.}
\vskip4mm
{\bf\noindent Probes at the LHC}
\vskip2mm
In the  main production channels,  $pp \to Z'\to \ell^+
\ell^-$ ($\ell=e,\mu$) one would be able to measure
the mass $M_{Z'}$,  the total width $\Gamma_{tot}$ and the total cross section
 $\sigma_{\ell \ell}$.
 The quantity
$\sigma_{\ell\ell}\Gamma_{tot}$ would in turn
 yield  information on  an overall strength  of the $Z'$  gauge coupling.
On the other hand there is a  need for probes which are sensitive to the {\it
relative strength} of $Z'$ couplings.
The nature of such probes can be classified according to the
type of  channel in which they can be measured.

\begin{itemize}
\item{(i) The forward-backward asymmetry \cite{LRR}} ,
\item{ (ii) the ratio of cross-sections in different rapidity
bins \cite{ACL} }
\end{itemize}
constitute distributions, {\it i.e.}, ``refinements'' in the main production
channel.\footnote{If the proton
polarization were available the corresponding asymmetries  would also
to be useful \cite{FT}\ .  In other two-fermion channels
measurements of the $\tau$ polarization
in the $pp\to Z'\to  \tau^+\tau^-$
channel \cite{AAC} ,  is also a useful probe.
Measurements of the cross section in $pp\to Z'\to   jet\  jet$
channel is the only  probe available for the left-handed quark
coupling \cite{CLIV}\ .
Recent studies \cite{RM,MOH}\ indicate that the measurement of
the cross-section in this channel  might be possible with
appropriate kinematic cuts, excellent
dijet mass resolution, and  detailed knowledge of the QCD backgrounds. }
The  forward-backward asymmetry   was long recognized as useful to
probe a particular combination of
quark and lepton couplings.
The rapidity ratio is a useful complementary probe separating the $Z'$
couplings to
the $u$ and $d$ quarks due to the  harder valence $u$-quark distribution
in the proton relative to the $d$-quark.

\begin{itemize}
\item{(iii)
Rare decays $Z'\to W\ell\nu_{\ell}$ \cite{RI,CLII}\ ,}
\item{(iv)
associated productions $pp\rightarrow Z' V$
with
$V=(Z,W)$\cite{CLIV}\  and $V=\gamma$ \cite{RII}}
\end{itemize}
 provide another set
of useful probes in the the four-fermion final state  channels.
These probes have  suppressed rates  compared to the two-fermion channels.
Rare decays turn out to have sizable statistics, however only the
modes   $Z'\to W\ell \nu_{\ell}$ \cite{CLII,PACO,HR}\  with appropriate
cuts  are useful\footnote{$Z'\to Z \ell^+\ell^-$
 does not significantly discriminate between models.}  without
large standard model and QCD backgrounds.
These modes probe a left-handed leptonic coupling.
On the other hand the  associated productions turn out to be relatively  clean
signals \cite{CLIV}\  with slightly smaller statistics than rare decays.
They probe  a particular combination of  the gauge couplings to quarks
and are thus complementary  to  rare decays.
\vskip4mm
{\bf\noindent Probes at the NLC}
\vskip2mm
 There the  cross sections  and corresponding asymmetries  in the
two-fermion  final state channels,  $e^+e^- \rightarrow f\bar f$,
 will be measured.
The analysis is based on the  following  probes:
\begin{equation}
\sigma ^{\ell},\   \ \ R^{had} = {{\sigma ^{had}}\over
{\sigma ^{\ell}}},\ \ \ A_{FB}^{\ell}.
\label{ee}
\end{equation}
In the case that  longitudinal polarization of the    electron
beam is available there are additional  probes:
\begin{equation}
A_{LR}^{\ell, had},\ \ A_{LR,FB}^{\ell}
\label{eep}
\end{equation}
Here $\sigma$, $A_{FB}$, $A_{LR}$ and $A_{LR,FB}$  refer to the
corresponding cross-sections, forward-backward asymmetries, left-right
asymmetries and left--right--forward--backward asymmetries, respectively.
 The index $\ell$
refers to all three leptonic channels (considering only
$s$-channel exchange for electrons) and $had$ to all
hadronic final states.
The above  quantities  distinguish among different
models\cite{DLRSV} ,  however,  they do not yield  information on all the
 $Z'$ couplings.

If  one assumes\footnote{Note that at the LEP an efficient tagging of the charm
and
bottom flavors was achieved \cite{LEP} .} . an efficient  heavy flavor
($c,b,t$)
tagging  there are the following  additional observables:
\begin{equation}
R^{f}={\sigma ^{f}\over \sigma^\ell} ,\ \ A^{f}_{FB} \ ; \ f=c,b,t\ \ ,
\label{eef}
\end{equation}
and with available polarization:
\begin{equation}
A^{f}_{LR}\ ,\ \ A^{f}_{LR,FB}\ ;\ f=c,b,t\ .
\label{eeft}
\end{equation}

These  additional probes   in turn allow for complete  determination of
the $Z'$ gauge  couplings to ordinary fermions.\footnote{ See Ref.
\cite{ACLIII}
 for
detailed discussions.}
\vskip4mm
{\bf\noindent Determination of gauge couplings  at the LHC}
\vskip2mm
 We assume the
c.m. energy $\sqrt s= 16$ TeV\footnote{For the  new projected c.m. energy
 $\sqrt{s} =14$ TeV, the cross section in the main production channel
 decreases  by $\sim$30\% and thus the
statistical error bars on the probes increase by  14\% .}
and integrated luminosity  ${\cal L}_{int}= 100$
fb$^{-1}$.
We consider only statistical uncertainties associated with the  probes (i-iv),
which yield sufficient qualitative information.
Realistic fits, which include updated structure functions, kinematic cuts,
and detector  acceptances are expected to give larger uncertainties for
the couplings.

We consider the following typical models:
$Z_\chi$  in $SO_{10}\rightarrow SU_5\times U_{1\chi}$,
$Z_\psi$ in $E_6\rightarrow SO_{10}\times U_{1\psi}$,
$Z_\eta=\sqrt{3/8}Z_\chi-\sqrt{5/8}Z_\psi$  in
superstring inspired models in which $E_6$ breaks directly to a rank 5
group,and
$Z_{LR}$ in LR symmetric models.  For conventions
in the neutral current interactions see
Ref. \cite{LL} .
In the following we assume family universality,
 neglect $Z-Z'$ mixing  and  assume $[Q',T_i]=0$,
which holds for $SU_2 \times U_1 \times U_1'$ and LR models.
 Here,  $Q'$ is the $Z'$ charge  and $T_i$ are the $SU_{2L}$ generators which
incidentally is satisfied by  the above models.

\begin{table*}
\begin{center}
\begin{tabular}{r|cccc}
&$\chi$&$\psi$&$\eta$&$LR$\\ \hline
$\gamma^\ell_L$&$0.9\pm 0.018$&$0.5\pm 0.03$&$0.2\pm 0.015$&
$0.36\pm 0.007$\\
$\gamma^q_L$&0.1&0.5&0.8&0.04\\
$\tilde{U}$&$1\pm 0.18$&$1\pm 0.27$&$1\pm 0.14$&$37\pm 8.3$\\
$\tilde{D}$ & $9\pm 0.61$ & $1\pm 0.41$ & $0.25\pm 0.29$ & $65\pm 14$
\end{tabular}
\caption{\protect\cite{ACL} Values of the couplings (5)
for the typical
 models. The statistical error bars indicate how well the coupling could be
measured at the LHC (c.m. energy $\protect\sqrt s = 16$ TeV and
integrated luminosity ${\cal L}_{int}=100\,\hbox{fb}^{-1}$) for $M_{Z'}=1$
TeV.}
\end{center}
\end{table*}
\begin{figure}[ht]
\caption{\protect\cite{ACLIII} 90\%  confidence level ($\Delta \chi^2=6.3$)
contours   for the $\chi$, $\psi$ and
$\eta$ models are plotted for $\tilde U$, versus  $\tilde D$, versus
$\gamma_L^\ell$. The input data are for $M_{Z'}=1$ TeV  at the LHC
($\protect\sqrt
s = 16$ TeV and  ${\cal L}_{int}=100\,\hbox{fb}^{-1}$)
and include statistical errors only.}
\end{figure}
\vskip8mm
The relevant quantities \cite{CLIV,ACL}\  to
 distinguish between different models are
the charges, $\hat{g}^u_{L2}=\hat{g}^d_{L2}\equiv\hat{g}^q_{L2}$,
$\hat{g}^u_{R2}$, $\hat{g}^d_{R2}$, $\hat{g}^\nu_{L2}=\hat{g}^e_{L2}
\equiv\hat{g}^\ell_{L2}$, and $\hat{g}^\ell_{R2}$, and the gauge
coupling strength $g_2$.
The signs of the charges will be hard to determine at hadron
colliders. Thus the following
 four ``normalized''
observables can be probed \cite{CLIV} :
\begin{eqnarray}
&\gamma_L^\ell={{(\hat{g}^\ell_{L2})^2}\over
{{(\hat{g}^\ell_{L2})^2+(\hat{g}^\ell_{R2})^2}}},\ \
\gamma_L^q={{(\hat{g}^q_{L2})^2}\over{{(\hat{g}^\ell_{L2})^2+(\hat{g}^\ell_{R2})^2}}},\nonumber \\
&\tilde{U}={{(\hat{g}^u_{R2})^2}\over {(\hat{g}^q_{L2})^2}},\ \ \ \ \ \ \ \ \
\ \ \ \ \ \ \
\tilde{D}={{(\hat{g}^d_{R2})^2}\over {(\hat{g}^q_{L2})^2}}\ .
\label{tild}
\end{eqnarray}
The values of these couplings for the typical models and the corresponding
 statistical uncertainties   for the $\gamma_L^\ell$,  $\tilde U$, and $\tilde
D$ couplings
are  given in  Table I.\footnote{ $\gamma_L^q$ could be
determined \cite{CLIV} by
measuring the branching ratio $B(Z'\to q\bar q)$.  See the footnote on the
previous page.}
In particular, $\gamma_L^\ell$ can be determined very well, primarily due to
the small statistical errors for the rare decay modes.
On the other hand the  quark couplings have larger uncertainties.
In  Figure 1 90\% confidence level  ($\Delta \chi^2=6.3$) contours\footnote{
The 90\%\ confidence level  contours for  projections
on the more familiar  two-dimensional
parameter subspaces correspond to $\Delta \chi^2=4.6$.}
are given in a three-dimensional plot for $\tilde U$ versus $\tilde D$
versus $\gamma_L^\ell$ for
the  $\eta$, $\psi$ and $\chi$ models (the $LR$ model
 is in a  different region of the parameter space).
  Note a clear separation between the   models.

 For $M_{Z'}\simeq 2$ TeV a reasonable
discrimination between models and determination of
the couplings may still be  possible, primarily from the forward-backward
asymmetry and the rapidity ratio. However, for
$M_{Z'}\simeq 3$  TeV there is little ability
to discriminate  between models.
\begin{table*}
\begin{center}
\begin{tabular}{r|cccc}
& $\chi$ & $\psi$ & $\eta$ & $LR$ \\ \hline
$P_V^\ell$ & $2.0\pm0.08\,(0.26)$ & $0.0\pm0.04\,(1.5)$ &
$-3.0\pm 0.5\,(1.1)$ & $-0.15\pm 0.018\,(0.072)$  \\
$P_L^q$ & $-0.5\pm 0.04\,(0.10)$ &  $0.5\pm0.10\,
(0.2)$ &  $2.0\pm0.3\,(1.1)$ & $-0.14\pm
 0.037\,(0.07)$
\\
$P_R^u$ & $-1.0\pm0.15\,(0.19)$ & $-1.0\pm0.11\,
(1.2)$ &  $-1.0\pm0.15\,(0.24)$ & $-6.0\pm1.4\,
(3.3)$\\
$P_R^d$ & $3.0\pm0.24\,(0.51)$ & $-1.0\pm0.21\,(2.8)$
& $0.5\pm0.09\,(0.48)$ & $8.0\pm1.9\,(4.1)$\\
$\epsilon_A$ & $0.071\pm0.005\,(0.018)$
& $0.121\pm0.017\,(0.02)$ &
 $0.012\pm0.003\,(0.009)$ & $0.255\pm0.016\,(0.018)$
\end{tabular}
\end{center}
\caption{\protect\cite{ACLIII}\ The  value of the couplings  for
typical models
and  statistical error-bars as determined from the probes (1-4)
 at   the NLC (c.m. energy $\protect\sqrt s = 500$ GeV and
integrated luminosity ${\cal L}_{int}=20\,\hbox{fb}^{-1}$).  $M_{Z'} = 1$ TeV.
100\%\  heavy flavor tagging efficiency  and 100\%\ longitudinal polarization
of the
electron beam  is assumed for the first set of error bars, while the
error bars in   parentheses are for the probes without
polarization.}
\end{table*}

\vskip4mm
{\bf\noindent Determination of gauge couplings  at the NLC}
\vskip2mm
We assume  the c.m. energy    $\sqrt s=500$ GeV and the integrated luminosity
 ${\cal L}_{int}=10$ fb$^{-1}$. For probes (1-4) we use the exact tree level
expressions and  assume   100\%  efficiency for the heavy
flavor tagging   (probes (3-4)) and  100\%
 longitudinal polarization  of the
initial electron beam for probes (2) and (4).
Only statistical uncertainties  are taken into account.
 If a new $Z'$ is known to exist, a realistic
fit should include  full radiative corrections,
 true experimental cuts and detector acceptances, which are expected to
increase the uncertainties.

Because the photon couplings are vector-like and the $\ell$ couplings
 to $Z$ have the property
$\hat g_{L1}^\ell\simeq -\hat g_{R1}^\ell$ it turns out that probes (1-4)
 single out the $Z'$ leptonic couplings primarily in the
combinations $
\hat g_{L2}^\ell\pm\hat g_{R2}^\ell$.  We  therefore
chose the ${\hat g^\ell_{L2} - \hat g^\ell_{R2}}$ combination, which turns out
to
be a convenient choice for the typical models used in the analysis,
 to  ``normalize'' the four charges in  the following way:
\begin{equation}
P_V^\ell ={{\hat g^\ell_{L2} + \hat g^\ell_{R2}}\over
{\hat g^\ell_{L2} - \hat g^\ell_{R2}}},
\ P_L^q = {{\hat g^q_{L2}}\over
{\hat g^\ell_{L2} - \hat g^\ell_{R2}}},
\ P_R^{u,d} =  {{\hat g^{u,d}_{R2}}\over
{\hat g^q_{L2}}}\ . \label{eec}
\end{equation}
In addition, the  probes (1-4) are
sensitive to the  following ratio of  an  overall gauge coupling strength
divided by the ``reduced'' $Z'$ propagator:
\begin{equation}
\epsilon_A= (\hat g_{L2}^\ell
- \hat g_{R2}^\ell)^2 {{g_2^2}\over{4\pi \alpha }}{
{s}\over{M^2_{Z'} - s}}\ . \label{eps}
\end{equation}
 Here $\alpha$ is the fine structure constant.
 Note that  couplings \ref{tild} probed by  the LHC, do not determine
 couplings \ref{eec} unambiguously.
In particular,  determination of $\gamma_L^\ell$, $
\tilde U$ and $\tilde D$ at the LHC
would yield  an eight-fold ambiguity   for the corresponding three
 couplings in \ref{eec}.

Statistical uncertainties  for couplings (6-7) are
given  in Table II.
The $Z'$  charges  can be determined to around $10-20\%$. Poor determination of
couplings for the $\eta$ model is related to the small value of $\epsilon_A$
in this case.  Note that
relative error bars are typically  by a factor of $\sim 2$ bigger than the
corresponding ones
at the LHC. Without  polarization
the  error bars  increase by a factor $2-10$, and thus yield
only marginal information about the  couplings.  The $\psi$ model has
 particularly  poorly determined couplings without polarization.
In Fig. 2,  90\%\ confidence level ($\Delta \chi^2=6.3$) contours  are given
in a three dimensional plot of $P_R^u$ versus $P_R^d$ versus $P_V^\ell$  for
the $\chi$, $\psi$ and $\eta$ models (the $LR$ model is in a  different region
of the parameter space).

In the case of  smaller, say, 25\%,   heavy flavor tagging
efficiency   and  in the  case that
the electron beam polarization is  reduced to, {say},
50\%,  the determination of the couplings is poorer, however still useful.
For  $M_{Z'}\sim 2$  TeV, the uncertainties for the couplings in the
typical models  are too large to discriminate between models.
\vskip4mm
{\bf\noindent Conclusions}
\vskip2mm
The analysis  demonstrates complementarity
between the NLC and  the LHC  colliders, which in conjunction allow for
determination of the $M_{Z'}$, an overall $Z'$ gauge coupling strength as well
as a  unique determination of   {\it all} the quark and lepton charges with
sufficiently small error bars, provided $M_{Z'}\lsim 1$ TeV.
\newpage
{\bf\noindent Acknowledgment}
\vskip2mm
 I would like to thank
F. del Aguila and P. Langacker for enjoyable collaboration.
 The work  was supported
by the Department of Energy Grant \#
DE--AC02--76--ERO--3071, and the
Texas National Research Commission Laboratory.

\end{document}